# A Machine Learning-Based Method for Identifying Critical Distance Relays for Transient Stability Studies

Ramin Vakili, *Student Member, IEEE*, and Mojdeh Khorsand, *Member, IEEE*

*Abstract*—Modeling protective relays is crucial for performing accurate stability studies as they play a critical role in defining the dynamic responses of power systems during disturbances. Nevertheless, due to the current limitations of stability software and the challenges of keeping track of the changes in the settings information of thousands of protective relays, modeling all the protective relays in bulk power systems is a challenging task. Distance relays are among the critical protection schemes, which are not properly modeled in current practices of stability studies. This paper proposes a machine learning-based method that uses the results of early-terminated stability studies to identify the critical distance relays required to be modeled in those studies. The algorithm used is the random forest (RF) classifier. GE positive sequence load flow analysis (PSLF) software is used to perform stability studies. The model is trained and tested on the Western Electricity Coordinating Council (WECC) system data representing the 2018 summer peak load under different operating conditions and topologies of the system. The results show the great performance of the method in identifying the critical distance relays. The results also show that only modeling the identified critical distance relays suffices to perform accurate stability studies.

*Index Terms*-- Distance relays, identifying critical protective relays, modeling protective relays in stability studies, power system protection, random forest classifier, relay misoperation, transient stability study.

## I. INTRODUCTION

The dynamic responses of the major assets in power systems including generators, loads, and control systems along with the responses of the protection schemes are the two main aspects that define power system behavior during major disturbances [1]. Post-analysis of many of the prior blackouts and major outages show that unforeseen relay misoperations played a significant role in leading the system toward these catastrophic events [2], [3]. In this regard, protection systems are identified by the North American Electric Reliability Corporation (NERC) as critical reliability assets in power systems [4]. Therefore, proper modeling of protection systems in transient stability studies is crucial for obtaining a realistic assessment of system behavior [5].

There are different types of protective relays in bulk power systems that are needed to be included in transient stability studies to achieve a realistic assessment of system behavior. Some of these protective relays, such as underfrequency load shedding (UFLS) and undervoltage load shedding (UVLS) relays, usually are included in transient stability studies. However, despite being one of the most common and critical protective relays in power systems, distance relays are usually not properly included in stability studies [5]. References [4]-[8] show that the results of transient stability studies performed without modeling distance relays might not capture the actual response of the system to a disturbance. This paper also shows the importance of modeling distance relays in stability studies with further case studies on the WECC system.

The most straightforward way to capture the behavior of distance relays in transient stability studies is to model all the relays in the system. However, modeling thousands of distance relays that exist in bulk power systems, such as the WECC system, is challenging due to two main reasons [6]:

- Commercial stability software tools, such as GE PSLF, have limitations on the number of dynamic models that can be included in the dynamic files used for stability studies. Therefore, modeling thousands of distance relays in a bulk power system overwhelms the dynamic file and exceeds the current limitation of the software.
- In bulk power systems, it is a challenging task to keep the updated setting information of thousands of distance relays in the dynamic file of the system since protection engineers change these settings for various purposes. The reach of the operation zones and the time delay for the operation of each zone are the two main settings that govern the functionality of distance relays. In an outdated dynamic file, the settings of one or more distance relays can be different from their actual values. It is shown in [6] that outdated relay settings might lead to an incorrect assessment of the dynamic behavior of power systems during major disturbances.

Due to these challenges, the need for a method to identify the critical distance relays for each type of contingency, i.e., the distance relays that are likely to operate during the contingency, has been identified by the industry [3], [5], and [9]. To address this need, various methods are proposed in the literature to identify the critical distance relays that are required to be included in transient stability studies.

A link between the power system simulator for engineering

(PSS/E) and the computer-aided protection engineering (CAPE) software—a software tool for analyzing the behavior of protective relays—has been developed in [10]. The link developed in [10] provides a platform for simultaneous assessment of protection system behavior and dynamic response of power systems during disturbances. During severe disturbances, the initiating event might have system-wide impacts and results in unstable power swings that can cause misoperations of several distance relays in the system at locations far from the initial fault location [9]. However, using the approach of [10], only the operations of the distance relays in the vicinity of initial fault locations are analyzed and the system-wide effects of disturbances on the distance relays are ignored. References [7], [11], and [12] have proposed different methods for identifying critical distance relays. The methods proposed in these references are based on the location and size of the initiating events. Therefore, these methods also fail to identify the distance relays which operate due to the system-wide effects of the initiating events.

As another method of identifying critical distance relays, the Independent System Operator (ISO) of New England monitors the impedance trajectories observed by relays in their planning studies. If any of the impedance trajectories traverse into the zone-3 reach of its related distance relay, which is considered 300% of line impedances for all the transmission lines in this initial study, the actual setting data of the relay is collected, and the relay is modeled in the planning studies [13]. The operation of distance relays affects the system behavior assessment for the remainder of the simulation. Therefore, after identifying the distance relay that operates first during the transient stability study, the results of the study might not properly reflect the actual behavior of the system for the remainder of the simulation. Hence, the rest of the critical distance relays identified by this method might be incorrect.

References [14] and [15] have proposed new methods for identifying the distance relays that are at the electrical center of a system and might misoperate due to unstable power swings during out-of-step (OOS) conditions. A generic method for identifying the critical distance relays for all types of contingencies is proposed in [5], which is based on solving an optimization problem using the prior generator grouping information and network structure. However, the methods proposed in [14], [15], and [5] only focus on the distance relay operations that occur due to unstable power swings. Thus, they do not consider the operations in the vicinity of the initiating event, which occur due to the initial impacts of the event. Moreover, the method in [5] provides a generic list of critical distance relays for all types of contingencies. However, the type and location of a contingency significantly affect the list of the critical distance relays that are required to be modeled for performing an accurate transient stability study. Therefore, providing a specific list of critical distance relays for each contingency is of great importance.

Different methods for designing distance relay schemes are proposed in [16]-[25] which enables these schemes to distinguish between a power swing and a fault condition. If these schemes are properly modeled in transient stability studies, they might be able to detect distance relay misoperations due to unstable power swings and OOS conditions. Nevertheless, none of these methods are able to identify all the critical distance relays which are required to be modeled for performing accurate transient stability studies.

The method proposed in [6] is an iterative algorithm that utilizes two methods of apparent impedance monitoring and minimum voltage evaluation to identify the critical distance relays for any contingency. Although this method can identify all the critical distance relays for a contingency, it imposes a heavy computational burden as several runs of transient stability studies should be performed in this method. This, in turn, leads to excessive simulation time.

A fast machine learning (ML)-based method is proposed in this paper that eliminates the drawbacks of the methods in the literature. The method trains an ML model which promptly identifies all the critical distance relays that need to be modeled in the transient stability study of a contingency. To train the model, the method utilizes the results of extensive offline transient stability studies of different types of contingencies under various operating conditions and topologies of the system. After being trained, the model uses the results of an initial early-terminated transient stability study of the contingency under study to identify the critical distance relays for that contingency. The number of critical distance relays identified by the method for any contingency is far less than the total number of distance relays in the system. Therefore, the method eliminates the challenges associated with modeling all the distance relays in the system. Moreover, it is indeed less challenging to keep the updated setting information for this small subset of distance relays in the dynamic files of bulk power systems. The major contributions of the paper can be summarized as follows:

- Unlike the methods proposed in the literature, the ML-based method developed in this paper is able to identify all the critical distance relays that are required to be modeled in the transient stability study of a contingency. This includes the distance relays that operate due to the initial impacts of the contingency and the ones that operate due to the system-wide impacts of the contingency.
- Unlike the other methods in the literature, the proposed method is very fast. Therefore, it can be used in planning studies, where a sheer number of contingencies are studied, to promptly identify the critical distance relays that need to be included in the transient stability studies of the contingencies under study.
- The proposed method is trained considering different operating conditions and topologies of the system under study. Therefore, as the results reveal, the method is robust against changes in the topology and the operating condition of the system and can yield great results even under a different topology and operating condition of the system.

The rest of this paper is organized as follows. Section II presents the proposed ML-based method for identifying critical distance relays. Section III describes the Random Forest algorithm, the metrics used to evaluate the performance of the trained RF model, as well as the grid search and the K-fold cross-validation methods. Section IV evaluates the performance of the trained model in terms of the metric used. Also, Section IV further assesses the performance of the method in identifying the critical distance relays and capturing the behavior of the system by performing transient stability studies of several contingencies on the WECC system using the proposed method. Conclusions are provided in Section V.

## II. IDENTIFICATION OF CRITICAL DISTANCE RELAYS

The ML-based method developed in this paper works based on the latent correlation between the impedance trajectories observed by distance relays during the early stages of a contingency and the behavior of the distance relays for several seconds later. Figure 1 shows the flowchart of the proposed method. During the training stage, to create a comprehensive dataset for training/testing an ML model, extensive transient stability studies are performed on different types of contingencies under different operating conditions and topologies of the system. The contingencies studied include bus faults and line faults followed by removing one or more transmission lines, as well as generator outage contingencies. During these studies, the real and imaginary parts of the impedance trajectories observed by the distance relays at the time of the fault and up to 1 cycle after the fault are captured and used as the features of the ML model. The operations of the distance relays during each contingency are used to label the dataset (1 for operation and 0 for not operation). Note that due to the software limitations, only the distance relays on the high voltage transmission lines with the voltage level of 345 kV and above are modeled in transient stability studies of the contingencies. If available, the results of the offline transient stability studies can be combined with the historical records of the impedance trajectories observed by distance relays in the system during prior contingencies along with the record of their operations. Note that after the model is trained using the comprehensive dataset, it is used to identify critical distance relays for any contingency. Hence, the process of building a comprehensive dataset through conducting extensive transient stability studies is only performed once at the training stage.

The problem of predicting distance relay operations is a binary classification problem (with two classes of 0 and 1). Throughout this paper, the following definitions for class 0 and class 1 of the classification problem hold:

- Class 0: This class represents the samples of no-operation of distance relays, i.e., the samples where a distance relay does not operate for the entire simulation time, in the dataset.
- Class 1: This class represents the samples of distance relay operations in the dataset.

Different types of ML algorithms can be used for classification problems. As it can train robust models with high accuracy and prediction/training speed, an RF classifier is used in this paper to train the ML model. To achieve the best performance, a two-stage grid search method is used to tune the hyperparameters of the RF model. During the grid search, to realistically evaluate the performances of the RF models trained with different combinations of hyperparameter values, the K-fold cross-validation method is utilized. The grid search and the

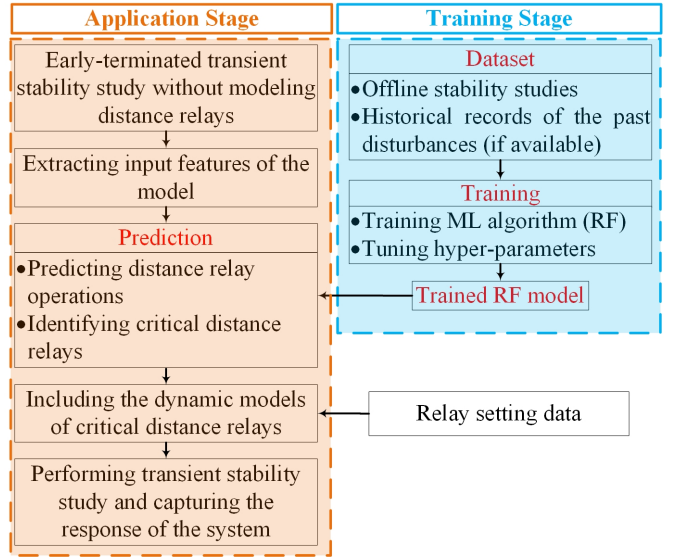

Fig. 1. The flowchart of the proposed method.

K-fold cross-validation methods are explained in Section III. C and Section III. D, respectively.

In the application stage, the trained model is used to identify the critical distance relays for any contingency under study. Note that this contingency might be a new contingency that has not been studied during building the dataset for training/testing the model; as the results of this paper confirm, the proposed approach performs well for such out-of-sample scenarios. First, the proposed method performs an initial transient stability study of the contingency under study without modeling any distance relays. Since the trained model only needs the impedance trajectories observed by distance relays up to 1 cycle after the fault as its input features, the initial transient stability study can be terminated 1 cycle after the fault, which can considerably reduce the simulation time. Note that the impedance trajectories observed by distance relays can be easily captured at each time interval of the transient stability study without any change in the existing practice of performing these studies and without any need to model the distance relays. Next, the method extracts the input features of the trained model from the results of the initial transient stability study. Then, the trained model predicts whether any of the distance relays in the system operate for the entire simulation time or not. The distance relays that are predicted to operate are identified as the critical distance relays for the contingency. Thus, at the next stage, their dynamic models are included in the dynamic file of the system. In this stage, the updated relay settings for the identified critical distance relays can be obtained (from the protection groups of utilities) and be included in the dynamic models of the distance relays to achieve a more precise assessment of the system behavior. Finally, using the stability software, the method performs a transient stability study with only modeling the critical distance relays to capture the response of the system to the contingency. Note that unlike the case of modeling all the distance relays, the small number of the critical distance relays identified by the proposed method can easily be included in stability studies without violating the current limitations of stability analysis software tools.

## III. THE RANDOM FOREST ALGORITHM, THE METRICS USED, THE GRID SEARCH METHOD, AND THE K-FOLD CROSS-VALIDATION METHOD

A brief explanation of the Random Forest algorithm along with its advantages, which make it suitable for predicting distance relay operations, are provided in this section. Also, the metrics, the grid search method, and the K-fold cross-validation method used in this paper are explained.

### A. Random Forest

The RF algorithm is a well-known and successful ML algorithm for classification and regression problems. Composed of multiple single decision trees (DTs), the RF operates as an ensemble ML algorithm. Each individual DT of the RF is trained on a subset of the dataset and uses a subset of the features as its input. In classification problems, the output of the RF is selected based on the most votes from its individual DTs [26]. Numerous advantages of the RF have made it a popular ML algorithm to be widely deployed in many problems [27]. Some of the most important advantages of the RF that make it suitable for the problem of this paper are as follows:

- One of the widely known problems of a single DT is overfitting. This causes DT to have a poor performance on new datasets. To overcome this problem, the RF algorithm trains multiple DTs on different parts of the training set using bootstrap aggregating or bagging techniques. This eliminates the overfitting problem of a single DT. Hence, the RF can have high accuracy on new datasets [28].
- The performance of the RF in handling unbalanced and non-linear datasets is outstanding [28]. For the application of this paper, as the dataset is unbalanced and has more cases of class 0 than class 1, this feature of the RF algorithm is of great importance.
- The RF has a high training and prediction speed, and it can easily handle high dimensional data, as it only uses a subset of features to train each DT. Also, as the RF algorithm is parallelizable, the process can be divided between multiple computers, which can significantly increase the speed of the algorithm. The high speed of the RF algorithm is favorable for the problem of this paper since it enables the RF to predict the operation of all the distance relays in bulk power systems in a short time [27], [28].

### B. Metrics

To evaluate the performance of the trained model two metrics of Recall and Precision are used in this paper. These metrics are introduced in this subsection. Note that, in the formulations of Recall and Precision, *TP* is the number of true-positive cases, i.e., the cases of distance relay operations (class 1) that the trained model correctly predicts as operation cases. *FN* is the number of false-negative cases, i.e., the distance relay operation cases that the trained model erroneously predicts as no-operation cases (class 0). Finally, *FP* is the number of false-positive cases, i.e., the no-operation cases of distance relays that the trained model erroneously predicts as operation cases.

Recall or true-positive rate (TPR) is the fraction of all the positive cases, i.e., distance relay operation cases in this paper, that the trained model correctly predicts as positive cases. Recall can be formulated using (1) [29].

$$Recall = \frac{TP}{TP+FN} \quad (1)$$

In the application of identifying critical distance relays, it is of the highest importance that the trained model does not miss any distance relay operation. In other words, ideally, the trained model should correctly predict all the distance relay operations. Hence, *FN* should be very low, and the Recall value should be very high (close to 1) to guarantee that all the distance relay operations are captured by the trained model. Therefore, in this paper, maximizing the Recall value is considered the objective of the grid search on the hyperparameters of the RF.

Precision is defined as the fraction of all the positive predictions of the trained model that are correct predictions. Precision can be formulated using (2) [29].

$$Precision = \frac{TP}{TP+FP} \quad (2)$$

In this paper, a high number of false-positive cases show that the trained model is prone to misclassify many of the no-operation cases of distance relays as operation cases and identify those distance relays as critical. Therefore, many distance relays might be included in the dynamic file of the system, which is undesirable; modeling many distance relays might cause the same challenges as modeling all the distance relays. However, modeling additional distance relays in the final stability analysis will not impact the accuracy of the analysis as long as the number of these relays does not exceed the limitation of transient stability analysis software tools and the relay settings are accurate. Thus, although maximizing the Precision value is not considered the main objective, it is tried to achieve a suitable value for Precision that does not result in the modeling of a large number of distance relays in transient stability studies.

### C. Grid search method

The grid search, also known as parameter sweep, is a traditional method to optimize the hyperparameters of an ML algorithm. In grid search, an exhaustive search is performed through a specified range of the hyperparameter space of the ML algorithm. A performance measure should be considered to direct the grid search in optimizing the hyperparameters. In this paper, as mentioned earlier, maximizing Recall value is considered the performance measure. The performance metric is measured by applying the cross-validation method at each iteration of the grid search method [30]-[32].

The grid search method can be easily used to optimize the hyperparameters of an ML model. Its biggest disadvantage is that when the number of hyperparameters to be tuned increases, this method can be computationally heavy and time-consuming. However, because the hyperparameters to be tuned are usually independent of each other, the processing task of the grid search can be divided into several parallel tasks with little or no effort, and each task can be performed with a separate computer [33].

### D. K-fold cross-validation

The K-fold cross-validation method is used to evaluate the performances of the models trained at each iteration of the grid search on new and out-of-sample cases. The method divides the dataset into *K* folds and at each iteration, it selects one of the folds as the test set and the remaining *K-1* folds as the training set. Then, a model is trained on the training set and is tested on the test set. The score/error of the model is recorded at each iteration. The iterations stop when all the folds are used as a test set. The average of the recorded scores/errors (Recall value in this paper) in all the iterations is the performance metric for comparing the performances of the trained models [34]-[35].

IV. CASE STUDIES

The performance of the proposed method is evaluated on the actual WECC system data representing the 2018 summer peak load case. The system includes 23297 buses, 18347 lines, 4224 generators, and 9050 transformers. The maximum generation capacity and the load of the system are 281.38 GW and 174.3 GW, respectively. The PSLF software is used for performing transient stability studies to create the dataset required for training and testing the RF model. Overall, 929 contingencies under different operating conditions and topologies of the system are studied. To be able to observe the behavior of distance relays in these studies, a reference dynamic file is created for the WECC system that includes two types of widely used distance relay models from the PSLF model library [36], namely, "*Zlin1*" and "*Zlinw*".

*Zlin1* is a distance relay model in the PSLF model library that includes three operation zones [36]. *Zlin1* needs to be modeled on each line with its specific settings. On the other hand, *Zlinw* is a generic distance relay model in the PSLF model library with two operation zones [36]. Unlike *Zlin1*, *Zlinw* does not need to be modeled for every line with its specific settings; rather, it monitors all the lines between its minimum and maximum voltage settings. Each of these distance relay models is used to model the distance relays on a subset of the lines of the system depending on the voltage level of the lines. Table I shows the voltage level of the lines whose distance relays are modeled with the *Zlin1* and *Zlinw* models.

Table I. The voltage level of the lines modeled with *Zlin1* and *Zlinw* models

| Model | Transmission lines voltage level |
|---|---|
| Zlin1 | > 345 kV |
| Zlinw | 100 kV – 345 kV |

Note that since the *Zlin1* model needs to be included for each line, separately, it is only modeled on the lines with a voltage level equal to or higher than 345 kV since these lines are the most critical lines of the system. Modeling *Zlin1* on every line overwhelms the dynamic file of the system and exceeds the current limitation of the PSLF software tool. Therefore, for the lines with a lower voltage level, the *Zlinw* model is used, as this model does not need to be included on each line, separately.

The well-known "scikit-learn" library [37] in Python is used to train the RF model. To achieve the best performance, a two-stage grid search method with the objective of maximizing the Recall metric is used to tune the hyperparameters of the RF model. Furthermore, the K-fold cross-validation method with 5 folds is used to assess the performances of the trained models.

The most important hyperparameters of the RF classifier are *class weights*, *the number of trees in the forest*, *the maximum depth of the trees*, *the minimum sample split (MSS)*, *the minimum sample leaf (MSL)*, *the number of features considered when looking for the best split*, and *whether bootstrap samples are used when building trees*. Precise tuning of these hyperparameters is critical for enhancing the performance of the trained model. Explaining the role of each of these hyperparameters in the performance of the RF model is out of the scope of this paper, and more details in this regard can be found in the "scikit-learn" library [37]. These hyperparameters can take a wide range of values. Hence, at the first stage of the grid search, an approximation of the best value for each hyperparameter is obtained. Then, at the second stage, the grid search method searches around the approximate value obtained from the first stage to find the exact value for each hyperparameter that yields the best performance.

To show the impact of changing each hyperparameter on the performances of the trained models, the change in the Recall value for the various *number of trees in the forest* and *the maximum depth of trees* is provided in Fig. 2(a). Also, the change in the Recall value for the various *class weights* and the *number of trees in the forest* is provided in Fig. 2(b). Note that, in Fig. 2(b), the set of the numbers on the X-axis (the *class weight* axis) shows the weight of each class. For example, "0:1, 1:10" means the weight of 1 for class 0 and the weight of 10 for class 1. The other sets of numbers are interpreted, likewise. The "balanced" weight on this axis means a weight for a class that is inversely proportional to the class frequency in the input dataset [37]. For example, if the "balanced" option is used for the *class weight* and the number of samples in the dataset belonging to class 1 is 3 times that belonging to class 0, the weight of class 1 will be 1/3 of the weight of class 0.

Fig. 2(a) shows that in comparison to the *number of trees in the forest*, the *maximum depth of trees* has slightly more impact on the performance of the trained models. It also shows that the trained model has a better performance when the *maximum depth of trees* is set to 10. Fig. 2(b) shows that the *class weights* have the highest impact on the performance of the trained model, and as the weight of class 1 increases, the Recall value increases, as well. Note that, this increase comes at the price of deteriorating the Precision value of the trained model. Therefore, although the weight of 100 for class 1 gives the highest Recall value in the first stage of the grid search method (above 0.99), the performance of the trained model in terms of the Precision value is very poor for this weight of class 1 (below 0.5). Thus, for the first stage, the weight of 10 is selected for class 1, which yields a high Recall value (around 0.979), while it keeps the Precision at a reasonable value (above 0.7).

Table II shows the best values obtained for each hyperparameter from the grid search method. Using the values provided in Table II for training the RF model yields the best results in terms of the Recall value (while maintaining a high Precision value). The performance of the trained model is evaluated on the entire dataset, including over 900 contingencies, using the K-fold cross-validation method. The results reveal that the trained model has Recall and Precision values of 0.981 and 0.737, respectively. This high Recall value of the trained model can ensure that it can correctly identify all the critical distance relays for a contingency and does not miss any distance relay operation, even under different operating points and topologies of the system. Whereas the reasonably high value of Precision shows that the model does not identify many distance relays in the system as critical. Using the hyperparameters obtained from the grid search, the final RF model is trained on the whole dataset and implemented in the proposed method to identify the critical distance relays for any contingency under study in the application stage.

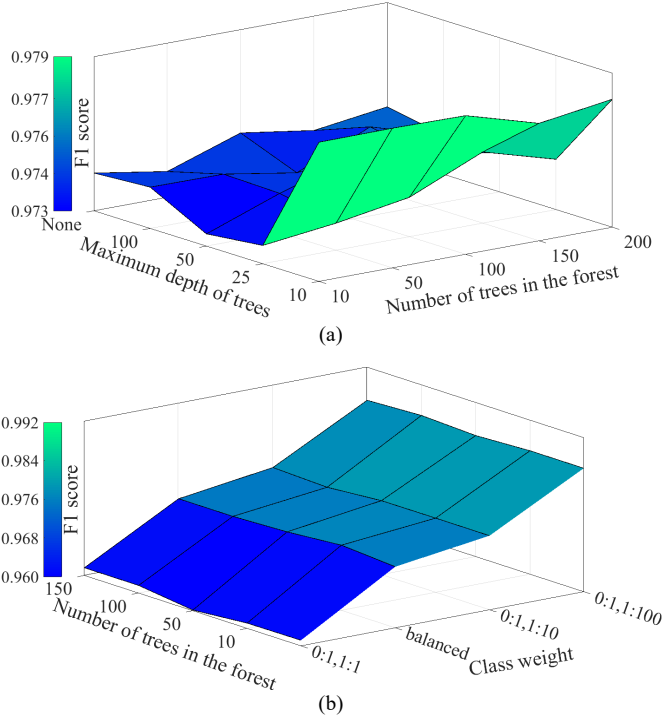

Fig. 2. The change in the Recall value for different values of the *maximum depth of trees* and the *number of trees in the forest* (a), as well as for the different values of *class weights* and the *number of trees in the forest* (b).

Table II. The selected hyperparameters of the RF model

| Hyperparameter | Tuned Value |
| --- | --- |
| Class weights | 1 for class 0 and 20 for class 1 |
| Number of trees in the forest | 15 |
| Maximum depth of trees | 15 |
| MSS | 2 |
| MSL | 2 |
| The number of features | "auto"—equal to the square root of the number of all the features |
| Bootstrap | "False"—meaning the whole dataset is used to build each tree |

To further illustrate the performance of the proposed method in identifying the critical distance relays and capturing the precise response of the system during various contingencies three different contingencies are considered for more detailed studies. Different operating conditions and topologies of the system are considered in these studies. For each contingency, three different cases are analyzed:

- Case 1: To show the impact of modeling distance relays on the response of the system, the first case considered in studying each contingency performs the transient stability studies without modeling distance relays.
- Case 2: In this case, the reference dynamic file, which was used for performing transient stability studies during creating the dataset stage, is used for performing transient stability studies of each contingency. Note that, as mentioned earlier, this dynamic file includes *Zlin1* models for the lines with a voltage level equal to or higher than 345 kV and a *Zlinw* model for the lines with a voltage level between 100 kV and 345 kV. This case is referred to as the reference case throughout this paper.
- Case 3: In this case, the proposed method is used to identify the critical distance relays for the contingencies under study. Then, the transient stability studies are performed with only including *Zlin1* models for the identified critical distance relays. Similar to the second case, the *Zlinw* model is used in the third case to monitor the lines with a voltage level between 100 and 345 kV. Note that, the proposed method can identify all the critical distance relays on the lines at any voltage level. However, to be able to compare the results of the method with the second case (the reference case), the settings of the method are modified to only identify the critical distance relays on the lines with voltage levels of 345 kV and above.

In both the second and the third cases, zone-1, -2, and -3 reaches of *Zlin1* models are considered to be 80%, 120%, and 220% of the related line impedances, respectively. Time delays of 0, 0.2, and 0.3 seconds are considered for zone-1, -2, and -3 operations of *Zlin1* models, respectively. The same zone-1 and -2 reaches and time delays are used for the *Zlinw* model. Circuit breaker delay time is set to 0.05 seconds, which means that it takes 0.05 seconds for the circuit breaker to open the line after receiving the tripping signal from its distance relay. Note that, although generic relay settings are considered for simplicity, the method can be used with precise relay settings obtained from the protection groups of electric utility companies to provide a more realistic assessment of system behavior.

Note that, these contingencies are only provided to show in more detail the performance of the proposed method and the response of the system for three simulated out-of-sample contingencies. However, the values of 0.981 and 0.737 reported earlier for the Recall and Precision metrics are obtained from testing the model on the entire dataset (over 900 contingencies) using the K-fold cross-validation method. To protect the proprietary data, arbitrary numbers are used throughout the paper to represent power system assets.

*Contingency 1*: In this contingency, a bus fault occurs on Bus 3 of the system and is cleared after 4 cycles by removing the three 500 kV transmission lines that comprise California Oregon Intertie (COI). During the summer peak load, COI transfer a total power of around 4,113 MW from the north to south of the WECC system. As these three lines transfer a considerably high amount of power, they are very critical tie lines of the WECC system, and their outage is a known critical emergency for the WECC system, which has the potential to jeopardize the system stability. Therefore, this critical *N-3* outage has been considered for a more detailed analysis. To evaluate the performance of the proposed method under new and unseen pre-fault operating conditions, a new pre-fault operation condition of the WECC system is considered for this contingency. In the new operating condition, a uniform increase of 2 percent in the large loads (above 100 MW) of the areas that import power through the COI, which leads to a 55.71 MW increase in the net load of these areas, is considered. This increase in the load is compensated by three of the largest generators existing in the areas that send power through COI, without violating the limitations of any generation unit in the system. The results of transient stability studies performed for Case 1 (modeling no distance relay), Case 2 (the reference case), and Case 3 (the proposed method) of this contingency are illustrated in Fig. 3 (a), (b), and (c), respectively.

Comparing Fig. 3 (a) and (b) shows that without modeling distance relays the dynamic response of the system in terms of the relative rotor angles of a set of generators is completely different from that of the reference case. Also, comparing Fig. 3 (b) and (c) illustrate that the proposed method can capture the behavior of the system exactly as the reference case, even for severe disturbances such as the disturbance of this contingency and a new pre-fault operating condition of the system. For the sake of clarity, only the rotor angles of a set of selected generators are shown. However, the relative rotor angles of all the generators show exactly the same response in the reference case and the proposed method.

In this contingency, with the distance relays modeled in the system, the network solution diverges at 6.321 (s) and the transient stability study terminates. The severe disturbance of COI outage has widespread effects on the system and leads to many distance relay operations. In the real-world system, proper remedial actions, such as load shedding and controlled islanding, are considered for this contingency. As in this paper, we do not have access to the remedial actions employed in the WECC system, transient stability studies are performed without considering the remedial actions. However, if the remedial actions are included, the proposed method is expected to be able to capture the behavior of the system.

Note that, improving the reliability and resiliency of power systems by designing proper preventive/remedial actions requires a detailed assessment of the responses of different assets and protection schemes in the system, and only analyzing if the system maintains its stability or not during a contingency is not sufficient. Thus, although in this contingency, the system becomes unstable in either case of modeling or not modeling distance relays, to have a precise assessment of the system response, which is crucial for devising proper preventive and remedial actions, modeling distance relays is necessary.

Table III shows the list of the relays that operate in the reference case (Case 2) and the proposed method (Case 3). Lines 1,2, and 3 are in the area that imports power through COI, and line 4 is in the area that exports power through COI (Lines 1-4 are in the vicinity of the fault location). Therefore, the distance relays of these lines operate due to the initial impacts of the disturbance. The ability to predict the operations of these distance relays shows that the method can correctly capture the distance relay operations due to the initial impacts of disturbances. All other lines listed in Table III are in the areas far from the fault location. Lines 8, 10, 15, and 17 are the tie-lines that connect two areas far from the fault location. The ability to correctly predict the operations of the distance relays of these lines and identify them as critical shows that the method can correctly capture the distance relay operations that occur due to the system-wide impacts of disturbances. Hence, it is observed that the method can identify all the critical distance relays for this severe contingency under the new pre-fault operating condition of the system.

*Contingency 2*: The same *N-3* COI outage that is analyzed in *Contingency 1* is also considered for *Contingency 2*. However, to evaluate the performance of the method in the case of a different pre-fault topology of the system, the pre-fault topology of the system is modified by taking two critical 500 kV transmission lines of the system out of service.

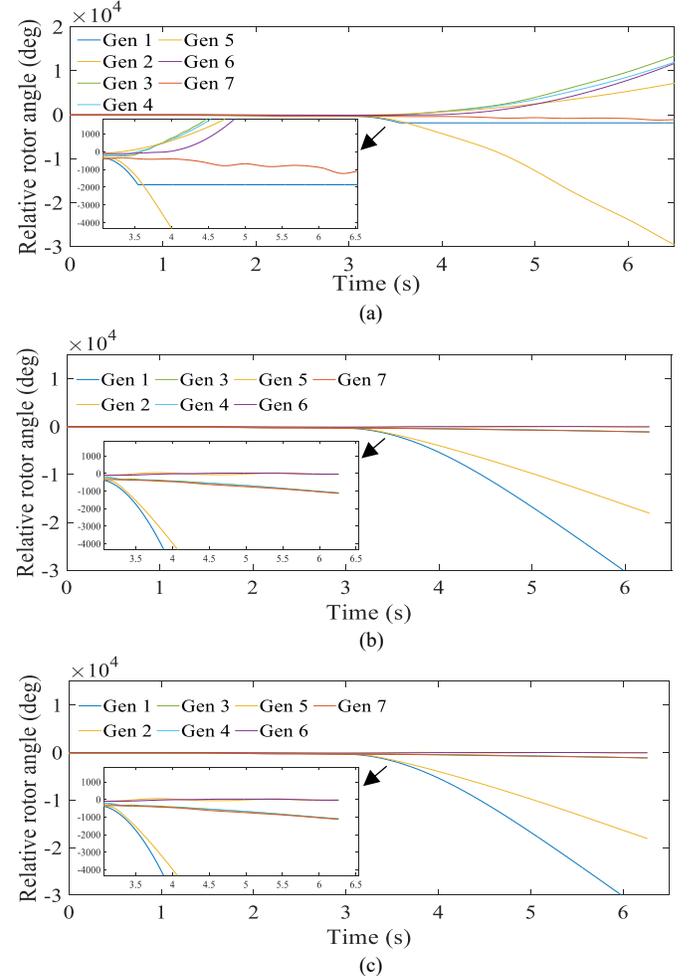

Fig. 3. The relative rotor angles of a set of generators in *Contingency 1*: (a) Case 1, (b) Case 2, and (c) Case 3.

Table III. The list of all distance relay operations in *Contingency 1*.

| The reference case | | The proposed method | |
| --- | --- | --- | --- |
| Time (s) | Relay | Time (s) | Relay |
| 1.050 | Lines 1, 2, 3 | 1.050 | Lines 1, 2, 3 |
| 1.888 | Line 4 | 1.888 | Line 4 |
| 2.629 | Line 5 | 2.629 | Line 5 |
| 2.671 | Line 6 | 2.671 | Line 6 |
| 2.683 | Lines 7 | 2.683 | Lines 7 |
| 2.767 | Line 8 | 2.767 | Line 8 |
| 2.771 | Line 9 | 2.771 | Line 9 |
| 2.950 | Line 10 | 2.950 | Line 10 |
| 2.979 | Line 11 | 2.979 | Line 11 |
| 3.009 | Line 12 | 3.009 | Line 12 |
| 3.025 | Line 13 | 3.025 | Line 13 |
| 3.179 | Line 14 | 3.179 | Line 14 |
| 5.921 | Line 15 | 5.921 | Line 15 |
| 6.321 | Line 16, 17 | 6.321 | Line 16, 17 |

The results of transient stability studies performed for Case1 (modeling no distance relay), Case 2 (the reference case), and Case 3 (the proposed method) of this contingency are illustrated in Fig. 4 (a), (b), and (c), respectively. As shown in Fig. 4 (a), without modeling distance relays, the response of the system in terms of the rotor angles of a set of generators is different from the reference case. Also, comparing Fig. 4 (b) and (c) demonstrates that the system response is similar in both the reference case and the proposed method, showing the accuracy of the method in capturing the response of the system. Similar to *Contingency 1*, in this contingency, with modeling distance relays, the network solution in the stability software diverges at 6.259 (s) and the transient stability study terminates.

Table IV provides the list of all distance relay operations in the reference case (Case 2) and the proposed method (Case3). Similar to *Contingency 1*, distance relay operations occur both on the lines close to the fault location and the lines far from it. As seen in Table IV, the method can correctly predict all the distance relay operations in this contingency, which shows the ability of the method to identify distance relay operations due to the initial and system-wide impacts of this severe contingency under the new pre-fault topology of the system.

Table IV. The list of all distance relay operations in *Contingency 2*.

| The reference case | | The proposed method | |
|---|---|---|---|
| Time (s) | Relay | Time (s) | Relay |
| 1.050 | Lines 1, 2, 3 | 1.050 | Lines 1, 2, 3 |
| 1.917 | Line 4 | 1.917 | Line 4 |
| 2.692 | Line 5 | 2.692 | Line 5 |
| 2.854 | Line 9 | 2.854 | Line 9 |
| 2.863 | Lines 6, 8 | 2.863 | Lines 6, 8 |
| 2.884 | Line 7 | 2.884 | Line 7 |
| 3.046 | Line 10 | 3.046 | Line 10 |
| 3.063 | Line 11 | 3.063 | Line 11 |
| 3.092 | Line 12 | 3.092 | Line 12 |
| 3.096 | Line 13 | 3.096 | Line 13 |
| 3.275 | Line 14 | 3.275 | Line 14 |
| 5.925 | Line 15 | 5.925 | Line 15 |
| 6.242 | Line 17 | 6.242 | Line 17 |
| 6.259 | Line 16 | 6.259 | Line 16 |

*Contingency 3*: To evaluate the performance of the proposed method in another type of disturbance, an *N-3* generator outage contingency is considered here. The selected generators are among the generators with the highest active power generation in the system. These three generators produce a total active power of around 1,268 MW. The pre-fault operating condition of *Contingency 1* is considered in this contingency, as well. This *N-3* generator outage has widespread impacts throughout the system, causing several distance relay misoperations due to unstable power swings. Therefore, identifying these critical distance relays by the proposed method further shows its ability to identify the distance relay operations due to unstable power swings at locations far from the location of the initial event.

The results of transient stability studies performed for Case 1 (modeling no distance relay), Case 2 (the reference case), and Case 3 (the proposed method) are shown in Fig. 5 (a), (b), and (c), respectively. Comparing Fig. 5 (a) and (b), it is revealed that without modeling distance relays, transient stability studies cannot capture the actual response of the system. Also, as seen in Fig. 5 (c), the response of the system in the case of using the proposed method is similar to that in Case 2 shown in Fig. 5 (b).

Table V provides the list of distance relay operations in both the reference case and the proposed method. As seen in this table, the method can correctly identify all the distance relays that operate in the reference case as critical. This further shows the capability of the method in predicting distance relay operation due to the system-wide impacts of a disturbance.

Analyzing these contingencies shows that the proposed method can correctly identify the critical distance relays and capture the precise response of the system for any type of contingency even under different operating conditions and topologies of the system. To further show that to what extent the method reduces the number of distance relay models included in transient stability studies, Table VI is provided. Table VI shows the number of distance relays identified as critical in each contingency. It also shows what percentage of the total number of distance relays in the reference case are identified as critical distance relays in each contingency. As seen in Table VI, the total number of critical distance relays in each contingency is less than 4.01% of the total number of distance relays in the reference case. Therefore, the dynamic models of this small number of critical distance relays with their updated settings can easily be obtained and included in the dynamic file of the system without violating the current limitation of stability software tools. This also significantly decreases the maintenance burden of keeping the distance relay setting information updated in the dynamic file of the system and the computational burden of the computing system.

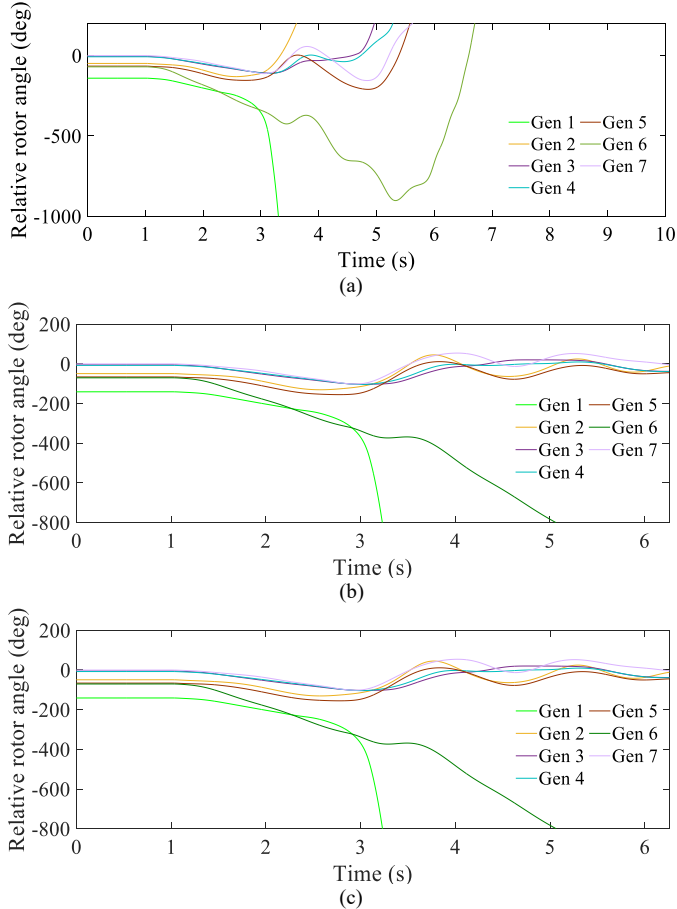

Fig. 4. The relative rotor angles of a set of generators in *Contingency* 2: (a) Case 1, (b) Case 2, and (c) Case 3.

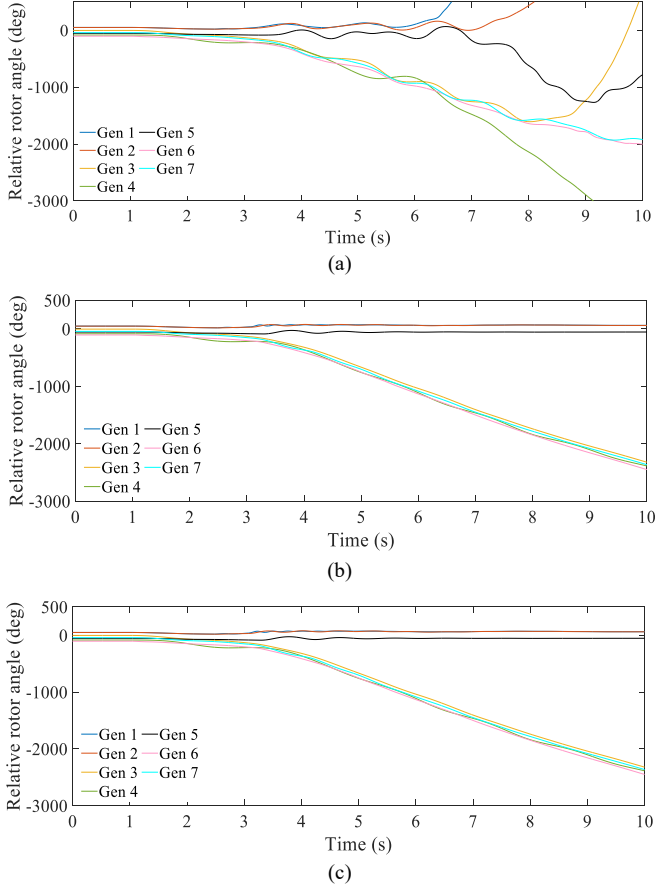

Fig. 5. The relative rotor angles of a set of generators in *Contingency 3*: (a) Case 1, (b) Case 2, and (c) Case 3.

Table V. The list of all distance relay operations in *Contingency 3*.

| The reference case | | The proposed method | |
|---|---|---|---|
| Time (s) | Relay | Time (s) | Relay |
| 3.075 | Lines 15 | 3.075 | Lines 15 |
| 3.283 | Line 17 | 3.283 | Line 17 |

Table VI. The total number of the identified critical distance relays.

| Contingency | Total number of the identified critical distance relays | Percentage of the total number of distance relays |
|---|---|---|
| 1 | 37 | 4.01% |
| 2 | 35 | 3.05% |
| 3 | 26 | 2.10% |

For the contingencies studied in this section, the processing time of the three major processes performed in the proposed method to identify the critical distance relays is provided in Table VII. These three major processes are as follows:
- Process 1: Performing the initial early-terminated transient stability analysis of the contingency under study.
- Process 2: Extracting the required features of the ML model from the results of the transient stability study and pre-processing them to be fed into the trained ML model.
- Process 3: Predicting the operation of all the distance relays in the system.

The processing time of performing transient stability studies depends on the type of the disturbance being studied as well as the operating condition and topology of the system. Therefore, as seen in Table VII, the processing time for performing the initial transient stability study is different for each contingency, whereas the processing time for extracting the features and the prediction time of the ML model are the same for all the contingencies. The total processing time for any contingency studied is less than 111 seconds. The three contingencies studied in this section are among the most severe disturbances of the WECC system. Hence, solving transient stability studies of these contingencies requires more processing time than other contingencies. The fact that the processing time of the method for these contingencies is less than 111 seconds guarantees that the processing time for any other contingency is also very small. This shows that the method can be used in planning studies to promptly identify the critical distance relays for a large number of the contingencies that are required to be studied.

Table VII. The processing time of the proposed method.

| Contingency | Process 1 (s) | Process 2 (s) | Process 3 (s) | Total (s) |
|---|---|---|---|---|
| 1 | 108.95 | 1.2 | 0.008 | 110.16 |
| 2 | 108.70 | 1.2 | 0.008 | 109.91 |
| 3 | 22.46 | 1.2 | 0.008 | 23.67 |

## V. Conclusion

This paper proposes an ML-based method to identify the critical distance relays required to be modeled for performing accurate transient stability studies of different types of contingencies. The method is based on training an ML model to learn the latent pattern between the impedance trajectories observed by distance relays during the early stages of a contingency and their operations for several seconds later. The RF classifier is used as the machine learning algorithm. After being trained using the dataset created from extensive offline transient stability studies and the records of historical outages, the RF model can predict the operation of distance relays during any contingency under study using the results of the early-terminated transient stability study of that contingency. The WECC system data representing the 2018 summer peak load is used as the test system. Using the K-fold cross-validation method, the performance of the trained model is evaluated on the entire dataset, and it is observed that the trained model has a great performance in terms of the Recall and Precision metrics. To show in more detail how the method can identify the critical distance relays for any contingency under study and capture the system response, three contingencies are studied under new topologies and operating conditions of the system.

The results show that the transient stability studies performed without modeling distance relays can manifest an inaccurate system behavior. Also, the results show the great performance of the proposed method in identifying the critical distance relays and capturing the precise behavior of the system. It is observed that in comparison to the reference case, the proposed method requires modeling a far smaller number of distance relays in transient stability studies (less than 4.01% of all the distance relays in the reference case for any of the contingencies studied). Also, it is observed that the total processing time for any of the studied contingencies is below 111 seconds, which illustrates the high speed of the method in identifying the critical distance relays in the system.

The small number of critical distance relays identified by the method can be easily modeled in stability studies without exceeding the limitations of stability software. Also, using this method, only critical distance relays are required to be accurately tracked for changes in their settings, which considerably reduces the maintenance burden. Finally, by reducing the number of distance relays modeled in the studies, the method significantly reduces the computational burden of the computing system. Thus, more dynamic models can be included in the analysis of various contingencies.

Similar strategies can be applied to identify other critical protective relays, such as generator protective relays, that play significant roles in defining the system response during a contingency and are required to be modeled in the transient stability study. Hence, developing similar methods for identifying the critical protective relays other than distance relays can be a proper continuation of this research.